\documentclass[12pt]{article}
\usepackage{epsfig}

\renewcommand{\theequation}{\arabic{section}.\arabic{equation}}
\renewcommand{\thefigure}{\arabic{section}.\arabic{figure}}
\newcommand{\lbl}[1]{\label{eq:#1}}

\newcommand{\vs}[1]{\rule[- #1 mm]{0mm}{#1 mm}}
\newskip\humongous \humongous=0pt plus 1000pt minus 1000pt

\newif\ifdtup

\newcommand{\eq}{\vs{2}\begin{equation}}
\newcommand{\be}{\vs{2}\begin{equation}}
\newcommand{\en}{\\[2mm]\end{equation}}
\newcommand{\bea}{\begin{eqnarray}}
\newcommand{\ena}{\end{eqnarray}}

\newcommand{\lapprox}{%
\mathrel{%
\setbox0=\hbox{$<$}
\raise0.6ex\copy0\kern-\wd0
\lower0.65ex\hbox{$\sim$}
}}
\newcommand{\gapprox}{%
\mathrel{%
\setbox0=\hbox{$>$}
\raise0.6ex\copy0\kern-\wd0
\lower0.65ex\hbox{$\sim$}
}}

\begin{document}
\bibliographystyle{plain}
\begin{titlepage}
\begin{flushright}
July 2001\\
\end{flushright}
\vspace{2.5cm}

\begin{center}
{\bf{\Large{\bf{Electromagnetic corrections to
$\gamma\pi^\pm\to\pi^0\pi^\pm$}}}}

\indent

{\bf Ll. Ametller$^a$, M. Knecht$^b$ and P. Talavera$^{b}$}
\\[0.5cm]
$^a$ {\it Dept. de F{\'\i}sica i Enginyeria Nuclear,
UPC\\
E-08034 Barcelona, Spain.}\\[.1cm]
$^b$ {\it Centre de Physique Th\'eorique, CNRS--Luminy, Case 907\\
F-13288 Marseille Cedex 9, France.}
\\[0.5cm]
\vfill
{\bf Abstract} \\
\end{center}
The amplitude for the anomalous transitions $\gamma\pi^\pm\to\pi^0\pi^\pm$ is 
analyzed within Chiral Perturbation Theory including 
electromagnetic interactions. The presence of a $t$-channel one-photon 
exchange contribution 
induces sizeable ${\cal{O}}(e^2)$ corrections which enhance the 
cross-section in the threshold region and bring 
the theoretical prediction into agreement with available data.
In the case of the crossed reaction $\gamma \pi^0 \rightarrow \pi^+ \pi^-$,
the same contribution appears in the s-channel and its effects are
small.

\noindent
\vfill
\begin{center}
{\bf Keywords:} \begin{minipage}[t]{9.5cm} Chiral Symmetry,
Chiral Perturbation Theory,\\ Scattering at low-energy,
Meson production.\end{minipage}
\end{center}
\end{titlepage}

\renewcommand{\thefigure}{\arabic{figure}}

\section{Introduction}
\renewcommand{\theequation}{\arabic{section}.\arabic{equation}}
\setcounter{equation}{0}
\label{intro}

The QCD matrix element of the 
electromagnetic current $j_{\mu}(x)$ between the 
vacuum and the three-pion state is described by a single invariant function 
$F^{3\pi}(s_+,s_-,s_0)$, {\it viz.} 
\eq
e\,\langle \Omega\,\vert j_{\mu}(0) \vert 
\,\pi^+(p_+)\,\pi^-(p_-)\,\pi^0(p_0) \rangle
\,=\,iF^{3\pi}(s_+,s_-,s_0)\,\epsilon_{\mu\nu\alpha\beta}
p_0^{\nu}p_+^{\alpha}p_-^{\beta}\,.
\en
In the absence of isospin breaking, $F^{3\pi}(s_+,s_-,s_0)$ is
symmetric with respect to permutations of the three variables,
with $s_{\pm}=(p_{\pm}+p_0)^2$, $s_0=(p_++p_-)^2$, and 
$s_++s_-+s_0=3M_{\pi}^2+q^2$, $q^2=(p_++p_-+p_0)^2$. Appropriate analytic 
continuations of $F^{3\pi}(s_+,s_-,s_0)$ at $q^2=0$
give the amplitudes $F^{3\pi}(s,\cos\theta)$ for the 
various $\gamma\pi\to\pi\pi$ pion photo-production processes, where $s$ and 
$\theta$ stand for the square of the total energy and for the scattering 
angle of the incoming pion in the center-of-mass frame, respectively. 

Standard PCAC techniques \cite{Terentev71,Adler71}
allow to relate this matrix element to the pion-photon-photon transition 
form factor $F^{\pi}(q_1^2,q_2^2)$ defined as
\eq
e^2\int d^4x e^{iq\cdot x}\langle \Omega\vert T\{j_{\mu}(x)j_{\nu}(0)\}\vert
\pi^0(p) \rangle \,=\, iF^{\pi}(q^2,(p-q)^2)\epsilon_{\mu\nu\alpha\beta}
q^{\alpha}p^{\beta}\,,
\en
with $F^{\pi}(q_1^2,q_2^2)=F^{\pi}(q_2^2,q_1^2)$.
This relation, which reads
\eq
\frac{F^{3\pi}(0,0,0)}{F^{\pi}(0,0)}\,=\,\frac{1}{eF_{\pi}^2}\,\big[
1 + {\cal O}(M_{\pi}^2) \big]\,,
\en
is exact in the chiral limit. In this limit, the value of each of the two 
amplitudes on the left-hand side of the above relation is actually known, 
since they both have their origin in the Wess-Zumino-Witten \cite{WZW}
anomalous contributions to the chiral Ward identities \cite{Aviv72}. 
In the case of $F^{\pi}(0,0)$, one has
\eq
F^{\pi}(0,0) \,=\,F^{\pi}_0\,
\big[
1 + {\cal O}(M_{\pi}^2) \big]\,,\quad F^{\pi}_0 \,=\, 
\frac{e^2N_c}{12\pi^2F_{\pi}}\,.
\label{Fpi0}
\en
Taking $F_{\pi}=92.4\pm 0.3$ MeV and the number of colours $N_c=3$ gives the 
value $F^{\pi}_0=0.0251\pm 0.0001$ GeV$^{-1}$, in good 
agreement with the experimental value 
$F^{\pi}_{\mbox{\scriptsize exp}}(0,0)=(0.025\pm0.001)$ GeV$^{-1}$ 
obtained 
from the $\pi^0\to\gamma\gamma$ width 
$\Gamma_{\pi^0\to\gamma\gamma}=7.74\pm 0.56$ eV \cite{PDG}, 
thus indicating that the chiral corrections in (\ref{Fpi0}) 
are very small. The analogous prediction for 
$F^{3\pi}(0,0,0)$ is 
\eq 
F^{3\pi}(0,0,0) \,=\,F^{3\pi}_0\,
\big[
1 + {\cal O}(M_{\pi}^2) \big]\,,
\quad\ F^{3\pi}_0\,=\,\frac{eN_c}{12\pi^2F_{\pi}^3}\,,
\label{F3pi0}
\en 
{\it i.e.} $F^{3\pi}(0,0,0)=(9.72\pm 0.09)$ 
GeV$^{-3}\big[1 + {\cal O}(M_{\pi}^2) \big]$, and it is likewise expected 
that the ${\cal O}(M_{\pi}^2)$ corrections are tiny. Experimental access to 
$F^{3\pi}(0,0,0)$ or $F^{3\pi}_0$ is however much less direct than in the 
case of $F^{\pi}(0,0)$. 

The presently most accurate experimental determination of the cross section 
for the reaction $\gamma (k) \pi^- (p_1)\rightarrow \pi^0 (p_0) \pi^- (p_2)$ 
proceeds through the Primakoff type pion pair production reaction of charged 
pions on nuclei
\eq
\label{reac}
\pi^-(p_1) + \mbox{(Z,A)} \to \pi^-(p_2) + \pi^0(p_0) + \mbox{(Z,A)}\,.
\en
In the Serpukhov experiment \cite{expe1}, the interaction is mediated by a 
virtual photon with momentum 
$q=p_2+p_0-p_1$, whose virtuality is small enough, 
$q_{\rm\mbox{\tiny max}}^2 \leq 2 \times 10^{-3}~\mbox{GeV}^{2}\ll M_\pi^2$,
so that it can be considered as a real photon. This experiment 
measured the total 
cross-section $\sigma^{\rm\mbox\tiny{total}}$ 
for the reaction (\ref{reac}) on different targets, and with
pion pairs  produced with a squared invariant mass $s=(p_2+p_0)^2$ up to 
$s_{\rm\mbox{\tiny max}} = 10 M_\pi^2$,
\eq
\sigma^{\rm\mbox\tiny{total}}_{\mbox{\scriptsize exp}}/{\mbox Z}^2 \,=\,
1.63\pm 0.23{\mbox{(stat.)}}\pm 0.13{\mbox{(syst.)}}\ {\mbox{nb}}\,.
\label{expresult}
\en
It is related to the cross-section $\sigma$ of the reaction 
$\gamma\pi^-\to\pi^0\pi^-$ through the equivalent 
photon approximation. Neglecting the $q^2$ dependence in 
$F^{3\pi}(s,t,u)$, with $t=(p_2-p_1)^2$, $u=(p_0-p_1)^2$, 
{\it i.e.} $F^{3\pi}(s,t,u)\approx F^{3\pi}(s,\cos\theta)$, 
$s+t+u = 3M_{\pi}^2$, one finds
\bea
\label{data}
\frac{\sigma^{\rm\mbox\tiny{total}}}{\mbox{Z}^2}
&=&
\frac{\alpha}{\pi}
\int_{(M_{\pi^0}+M_{\pi^\pm})^2}^{s_{\rm\mbox{\tiny max}}}
ds\, \frac{1}{s-M_{\pi^\pm}^2}\,
\left[\ln\left(
\frac{q_{\rm\mbox{\tiny max}}^2}{q_{\rm\mbox{\tiny min}}^2}\right)+
\frac{q_{\rm\mbox{\tiny min}}^2}{q_{\rm\mbox{\tiny max}}^2}-1\right]
\times \sigma\,,
\\
\label{cross}
\nonumber\\
\sigma &=&
\frac{1}{1024\pi}\,
\frac{\lambda^{3/2}(s,M_{\pi^\pm}^2,M_{\pi^0}^2)}{s^2}\,
(s-M_{\pi^\pm}^2)
\int_0^\pi d\theta \sin^3\theta 
\vert F^{3\pi}(s,\cos\theta)\vert^2\,.
\ena
Here, 
\[
q^2_{\rm\mbox\tiny{min}} = \left(\frac{s-M_{\pi^\pm}^2}{2E}\right)^2\,,
\]
with $E=  40$ GeV  being the energy of the incident pion beam, and 
$\lambda(x,y,z)=x^2+y^2+z^2-2xy -2xz - 2yz$.
The expressions (\ref{data}) and 
(\ref{cross})  were then fitted to the experimental value 
(\ref{expresult}), using however, in the 
ranges of $\theta$ and $s$ covered by the experiment, a constant average 
amplitude ${\overline F}^{\,3\pi}$, with the outcome
\eq
{\overline F}^{\,3\pi}_{\mbox{\scriptsize exp}}\,=\,
12.9\pm 0.9\pm 0.5\ {\mbox{GeV}}^{-3}  \,.
\label{F3pibar}
\en
This value of ${\overline F}^{\,3\pi}_{\mbox{\scriptsize exp}}$ has often 
been compared to $F^{3\pi}_0$, although the two quantities can in principle 
be quite different.

Since the point $s=t=u=0$ is unphysical, resort to theory is necessary in 
order to bridge the gap between $F^{3\pi}(0,0,0)$ and the amplitude 
$F^{3\pi}(s,\cos\theta)$, and thus establish the link between 
$\sigma^{\rm\mbox\tiny{total}}$ and $F^{3\pi}_0$. 
In the region where the Mandelstam variables are 
small as compared to the typical hadronic scale set by, say, the rho meson 
mass, a systematic expansion of the amplitude $F^{3\pi}(s,t,u)$ can be 
constructed within the framework of Chiral Perturbation Theory (ChPT)
\cite{Weinberg79,gasser},
\eq
F^{3\pi}(s,\cos\theta)\,=\,F^{3\pi}_0
\bigg[f^{(0)}(s,\cos\theta)+f^{(1)}(s,\cos\theta)
+f^{(2)}(s,\cos\theta)+\cdots\bigg]\,.
\label{Fchpt}
\en
In this notation, the low-energy theorems discussed above amount to 
$f^{(0)}(s,\cos\theta) = 1$ if isospin symmetry is preserved.
This chiral expansion has been performed to one loop some time ago
in Ref.~\cite{BBC} and, more recently, the two loop contribution
$f^{(2)}(s,\cos\theta)$ has become available as well~\cite{Hannah}. 
As expected for an $SU(2)_L\times SU(2)_R$ expansion, the corrections are 
indeed small in the threshold region, and  inserting the two-loop result 
into (\ref{cross}) leads to (for details, see~\cite{Hannah}) 
$\sigma^{\rm\mbox\tiny{total}}_{\rm\mbox\tiny{2loop}}/{\mbox{Z}^2}
=1.18$ nb. If instead
one keeps $F^{3\pi}_0$ as a free normalization constant in the
two-loop expression, then
the datum (\ref{expresult}) leads to the determination~\cite{Hannah}
\eq
F^{3\pi}_{0\,,{\mbox{\scriptsize exp}}}\,=\,11.4\pm1.3\,.
\label{F3piexp}
\en
This value is somewhat lower than (\ref{F3pibar}), and the discrepancy with 
the theoretical value quoted earlier is thus at the level of 1.3$\sigma$ 
only.

Still, one might wonder about the origin of this difference, since it is 
quite unlikely that it can be ascribed to yet higher order chiral corrections. 
This point of view is supported by the analysis of Ref. \cite{Hannah}, where 
the two-loop ChPT calculation was also supplemented by a dispersive 
approach, which captures at least part of the higher order ChPT 
contributions, but does not affect the result  (\ref{F3piexp}).
The amplitude $F^{3\pi}(s,\theta)$ has also been considered within different 
approaches, constituent quark models \cite{QL}, vector meson dominance models  
\cite{VMD1,VMD2}, and dispersively improved  vector meson dominance models  
\cite{dispVMD,Truong}. Although the model dependence in these studies is 
sometimes hard to quantify, the general tendency is towards producing values 
of the cross-section at low energies which are smaller than the experimental 
value if the normalization is kept fixed at $F^{3\pi}(0,0,0)=F^{3\pi}_0$.

In all the studies quoted so far, isospin breaking effects have 
not been taken into account. 
Their discussion is the purpose of the present work. Since the $m_d-m_u$ 
quark mass difference enters only at the level of ${\cal O}((m_d-m_u)^2)$ 
or ${\cal O}(e^2(m_d-m_u))$ effects, isospin violation in the 
$\gamma\pi\to\pi\pi$ reactions is essentially of electromagnetic origin.
The analysis of radiative corrections in the case of the low-energy
$\pi^0\pi^0\to \pi^+\pi^-$ amplitude 
\cite{marc_res}, would suggest that electromagnetic contributions are at most 
comparable to the two-loop effects in the threshold region. The situation is 
however qualitatively different in the case of the $\gamma\pi^\pm\to\pi^0\pi^\pm$ 
amplitude, due to the contribution arising from the exchange of a single 
photon between the $\gamma -\pi^0$ pair and the charged pion pair.
This already modifies the lowest order term, which becomes $f_0(s,\cos\theta) = 
1-2e^2 F_\pi^2/t$. 
As we shall see, this pole in the $t$-channel of the reaction 
$\gamma\pi^\pm\to\pi^0\pi^\pm$ is sufficiently close to the physical region 
in order to affect the amplitude in a sizeable way at low energies and bring
the theoretical and experimental values of $\overline{F}^{3\pi}$ into
agreement. In the case of the crossed reaction $\gamma \pi^0 \rightarrow \pi^+
\pi^-$, this pole occurs in the s-channel and its effects on the amplitude
are much less important.

The outline of the paper is the following:
in section \ref{em} 
we compute  the electromagnetic corrections to $f^{(0)}(s,\cos\theta)$ 
(tree level) and to $f^{(1)}(s,\cos\theta)$ for the process 
$\gamma\pi^\pm\to\pi^0\pi^\pm$.
The various counterterms involved 
are estimated in section \ref{results}, which is devoted to
the numerical analysis of our results.
Our conclusions are presented in section \ref{conclusions}.

\section{Radiative  corrections to $F^{3\pi}(s,\cos\theta)$ in ChPT}
\label{em}
\renewcommand{\theequation}{\arabic{section}.\arabic{equation}}
\setcounter{equation}{0}

In this section, we discuss electromagnetic effects in 
$f^{(0)}(s,\cos\theta)$ 
and in $f^{(1)}(s,\cos\theta)$ for the reaction 
$\gamma\pi^\pm\to\pi^0\pi^\pm$, 
which is the channel of interest for the Serpukhov experiment \cite{expe1}, 
but also for forthcoming experiments 
\cite{cebaf,compass,selex}. 
To this end, we use the formalism of ChPT 
in the situation where virtual photons are also present 
\cite{res,helmut}. In this context, the chiral counting is extended by 
considering the electric charge $e$ as a quantity of order 
${\cal O}({\mbox{\rm p}})$.
Actually, we shall use the two-flavour version of the 
formalism, discussed in Refs. \cite{marc_res} and \cite{ulf} to order one 
loop. As far as notation is concerned, we follow the first of these two last
references.

\subsection{Virtual photons in $f^{(0)}(s,\cos\theta)$}

Besides the ${\cal O}({\mbox{\rm p}}^2)$ mesonic and Maxwell terms, 
the lowest order 
chiral Lagrangian ${\cal L}^+_{(2)}$ in the even intrinsic parity sector
contains now ${\cal O}(e{\mbox{\rm p}})$ terms, arising 
from the minimal coupling to the photon field $A_\mu$, and a single 
${\cal O}(e^2)$ contact term described by a low-energy constant $C$, 
\bea
{\cal L}^+_{(2)} = {{F^2}\over 4}\,
\langle\,D^{\mu}U^\dagger D_{\mu}U\,+\,
\chi^\dagger U+U^\dagger \chi\,\rangle
\,-{1\over 4}F^{\mu\nu}F_{\mu\nu}\,
-\,{1\over{2\xi}}\,({\partial}\cdot A)^2\,
+\,C\,\langle\,Q U Q U^\dagger\,\rangle\ .
\label{p2}
\ena
The covariant derivative acting on $X$ is defined as usual,
$D_\mu X = \partial_\mu X -i r_\mu X + i X l_\mu$,
and in our particular case, we may take
$r_\mu = l_\mu =  A_\mu Q\,e\,, \quad \chi = 2 B {\cal M}$. 
Here ${\cal M}=\mbox{diag}(m_u,m_d)$ refers to 
the quark-mass matrix and $Q=\mbox{diag}(2/3,-1/3)$ denotes the quark charge 
matrix. {F}urthermore,  
$F_{\mu\nu}$ stands for the electromagnetic field strength tensor and 
hereafter we shall work in the 
Feynman gauge, $\xi=1$. The unitary matrix $U(x)$ is a 
parametrization
of the Goldstone boson fields, that may be taken as
\eq
\label{U}
U = e^{i\phi/F}, \quad
\phi=
\left( \begin{array}{cc}
 \pi^0   &  \sqrt{2}\pi^+\\
 \sqrt{2}\pi^- & - \pi^0 \\
    \end{array} \right)\,,
\en
although final results for observable quantities do not depend on this 
specific choice. The low-energy constant
$C$ gives the electromagnetic contribution to the charged pion mass,
\bea
&M_{\pi^0}^2 &= B (m_u+m_d)\ ,
\nonumber\\
&M_{\pi^\pm}^2 &= B (m_u+m_d)\,+\,2C {{e^2}\over{F^2}}\ ,\lbl{leadmass}
\ena  
which, for $F=F_\pi=92.4$~GeV, yields
\eq
\label{Z}
Z\ \equiv\ {C\over{F^4}}\ ={{M_{\pi^\pm}^2-M_{\pi^0}^2}\over{2e^2F^2}}
\approx 0.8\ .\lbl{leadZ}
\en

Anomalous processes involve contributions from the odd intrinsic parity sector,
which, at lowest order, is described by the Wess-Zumino-Witten 
Lagrangian \cite{WZW}. For processes involving photons, it reads
\begin{eqnarray}&&{\cal L}_{(4)}^-
=- \frac{N_c}{48 \pi^2} \epsilon^{\mu\nu\alpha\beta} \Bigl\{
A_\mu \langle Q ( \partial_\nu U \partial_\alpha U^\dagger
\partial_\beta U U^\dagger-
\partial_\nu U^\dagger \partial_\alpha U \partial_\beta
U^\dagger U) \rangle \nonumber\\
&&+4i \partial_\mu A_\nu A_\alpha
\langle Q^2 \partial_\beta U U^\dagger
+Q^2  U^\dagger \partial_\beta U - {1\over 2}Q U Q
\partial_\beta  U^\dagger + {1\over 2} Q
U^\dagger Q  \partial_\beta  U \rangle \Bigr\}\,.
\label{wzw}
\end{eqnarray}
The first term on the right-hand side of Eq.~(\ref{wzw})
is the relevant piece for the $\gamma \to
3 \pi$ transition at lowest order, and corresponds to the value 
$f^{(0)}(s,\cos\theta)=1$ in the absence of other contributions at lowest 
order. The second term in ${\cal L}_{(4)}^-$ is responsible for the 
$\pi^0 \to\gamma \gamma$ decay. If contributions involving virtual photons 
are considered, it also generates a contribution to $f^{(0)}(s,\cos\theta)$ 
arising from the one-photon exchange diagram of Fig. 1, where the 
$\pi^0$-$\gamma$-$\gamma$ vertex is reduced to its lowest order value 
$F^{\pi}_0$, and the electromagnetic form factor $F_V(t)$ to unity. 

In the case of the reaction $\gamma\pi^\pm\to\pi^0\pi^\pm$, the full 
expression of $f^{(0)}(s,\cos\theta)$ thus reads
\eq
f^{(0)}(s,\cos\theta)\,=\,1\,-\,2e^2\frac{F_{\pi}^2}{t}
\en
with
\eq\lbl{tu}
t = 2 M_{\pi^\pm}^2 -\frac{(s+M_{\pi^\pm}^2) 
(s+M_{\pi^\pm}^2 -M_{\pi^0}^2)  }{2s}
+\frac{(s-M_{\pi^\pm}^2)\lambda^{1/2}(s,M_{\pi^\pm}^2,M_{\pi^0}^2)}{2s} 
\cos\theta\,.
\en
Although the value $t=0$ is excluded from the physical region of the reaction 
$\gamma\pi^\pm\to\pi^0\pi^\pm$, small values of $t$ are possible, 
and actually not only  in the threshold region. 
Neglecting the pion mass difference for the time being, 
the threshold value is 
$f^{(0)}(4M_{\pi^\pm}^2,\cos\theta)=1+(\alpha/\pi)(4\pi F_{\pi}/M_{\pi^\pm})^2
=1.16$. Furthermore, in the forward direction, $\vert t\vert$ decreases as 
$s$ grows. The behaviour of $f^{(0)}(s,\cos\theta)$ in the range of $s$ covered
by the Serpukhov experiment \cite{expe1} is shown in Fig. \ref{fig.LO}. The 
increase in $f^{(0)}(s,\cos\theta)$, as compared to the constant value
$f^{(0)}(s,\cos\theta)=1$ corresponding to the absence of isospin breaking,
stays substantial for $\cos\theta\gapprox 0.8$ even away from threshold, 
and by itself increases the total cross-section 
${\sigma^{\rm\mbox\tiny{total}}}/{\mbox{Z}^2}$ by 16\%, 
from 0.92 nb to 1.07 nb.
In order to establish the robustness of this result, we next compute the 
radiative corrections to $f^{(1)}(s,\cos\theta)$ as well.

\subsection{Electromagnetic corrections in $f^{(1)}(s,\cos\theta)$}

The evaluation of the next-to-leading contribution 
$f^{(1)}(s,\cos\theta)$ involves loop diagrams with exactly one vertex 
from ${\cal L}_{(4)}^-$, and tree graphs, which involve the various
counterterms. In the even intrinsic sector, the counterterms consist of
the strong interaction low-energy constants $l_i$ of Ref.~\cite{gasser}, 
and of the electromagnetic constants $k_i$ \cite{marc_res,ulf}, 
corresponding to the decomposition 
(there are also ${\cal O}(e^4)$ counterterms,
but we shall not consider radiative corrections of this order)
\eq
\label{lagp4}
{\cal L}^+_{(4)} = {\cal L}^+_{\mbox{\tiny p}^4} 
+{\cal L}^+_{e^2\mbox{\tiny p}^2}\,.
\en
In the odd intrinsic parity sector, the next-to-leading order Lagrangian has 
a similar decomposition
\eq
\label{lagp6}
{\cal L}^-_{(6)} = {\cal L}^-_{\mbox{ \tiny e p}^5} 
+{\cal L}^-_{\mbox{\tiny e}^3 \mbox{\tiny p}^3 }\,.
\en
The expression of ${\cal L}^-_{\mbox{\tiny e p}^5}$, 
which involves a set of counterterms $A_i$,  has been worked out in 
Ref.~\cite{fearing},
and we do not reproduce it here for the sake of brevity.
The remaining term ${\cal L}^-_{ \mbox{\tiny e}^3 \mbox{\tiny p}^3}$
contains the  ${\cal O}(e^2)$ electromagnetic counterterms
contributing to the anomalous sector. To our knowledge, they have not been 
classified so far. 
For the time being, we shall collect the full contribution 
of all electromagnetic counterterms in the anomalous sector in an ``effective" 
low-energy constant $B_{\mbox{\tiny eff}}$.

In computing the loop graphs, we shall encounter ultraviolet divergences. 
These will be regularized within the same dimensional regularization scheme 
as used in Ref.~\cite{gasser}. The elimination of the divergences proceeds 
through the renormalization of the counterterms. The renormalized low-energy 
constants $l_i^r$, $k_i^r$, $A_i^r$ and $B_{\mbox{\tiny eff}}^r$ 
then depend on the renormalization scale $\mu$. As far as the 
contributions from the low-energy constants $l_i$ and $k_i$ are concerned, we 
express them in terms of the scale invariant quantities ${\bar l}_i$ and 
${\bar k}_i$, as defined in Refs.~\cite{gasser} and \cite{marc_res}, 
respectively. Of course, the final expression of $f^{(1)}(s,\cos\theta)$ 
has to be $\mu$-independent. We shall briefly address this issue at the end of 
this section. 

In order to present our results we have found it convenient 
to split the expression of $f^{(1)}(s,\cos\theta)$ into three components,
\eq
f^{(1)}(s,\cos\theta)\,=\,f^{(1)}_{\mbox{\tiny QCD}}(s,\cos\theta)\,+\,
f^{(1)}_{\mbox{\tiny FF}}(s,\cos\theta)\,+\,
f^{(1)}_{\mbox{\tiny IRD}}(s,\cos\theta)\,,
\en
which we shall describe in turn.
The contribution $f^{(1)}_{\mbox{\tiny QCD}}(s,\cos\theta)$ contains the 
diagrams with only mesons in the loop, as well as 
the counterterms $A_i$. 
It amounts to the calculation of Ref.~\cite{BBC}, except that one has to 
include effects of the the pion mass difference in the loops,
\bea
\hspace{-0.7cm}
f^{(1)}_{\mbox{\tiny QCD}}(s,\cos\theta)
& =&
- \Big\{ 8 M_{\pi^0}^2 \Big(2 A_{12}^r(\mu)-A_{13}^r(\mu) 
- A_7^r(\mu) + A_8^r(\mu)\Big) 
\nonumber\\&&
-\frac{16}{3}\Delta_{\pi}
\Big(A_7^r(\mu)-A_8^r(\mu)\Big) \Big\}
- \frac{1}{96 \pi^2 F_\pi^2}
\Big\{M_{\pi^0}^2+\,\frac{2}{3}\,\Delta_{\pi}
\nonumber\\&&
+(t+5M_{\pi^0}^2+2\Delta_{\pi})
(\,
\frac{M_{\pi^0}^2}{\Delta_{\pi}}\,\ln\frac{M_{\pi^\pm}^2}{M_{\pi^0}}-1)+
3M_{\pi^0}^2\ln\frac{M_{\pi^0}^2}{\mu^2}\,\Big\}
\nonumber\\&&
+\frac{1}{6F_\pi^2} {\bar J}_{+-}(t) (t-4M_{\pi^\pm}^2)
\nonumber\\&&
+\frac{1}{6F_\pi^2} {\bar J}_{+0}(u) 
\Big\{u-2(M_{\pi^\pm}^2+M_{\pi^0}^2)+\frac{\Delta_{\pi}^2}{u}
\Big\}
\nonumber\\&&
+\frac{1}{6F_\pi^2} {\bar J}_{+0}(s) 
\Big\{s-2(M_{\pi^\pm}^2+M_{\pi^0}^2)+\frac{\Delta_{\pi}^2}{s}
\Big\}
\,,
\label{f1QCD}
\ena
where $\Delta_{\pi}=M_{\pi^\pm}^2-M_{\pi^0}^2$.
$ {\bar J}_{PQ}(s) $ is the scalar two-point
function subtracted at $s=0$ \cite{gasser} (the subscript identifies the 
charges, and hence the masses, of the two pions in the loop),
and the terms $A_i^r(\mu)$ are the renormalized, 
scale-dependent counterterms 
from the anomalous ${\cal L}^-_{\rm \mbox{\tiny p}^6}$ Lagrangian. 
In practice, we shall only be interested in ${\cal O}(e^2)$ corrections, 
neglecting contributions involving higher powers of $e^2$. Terms like 
$\Delta_{\pi}^2/s$ or $\Delta_{\pi}^2/u$ can therefore be omitted from the 
expression (\ref{f1QCD}).

\begin{figure}[t]
\begin{center}
\epsfig{file=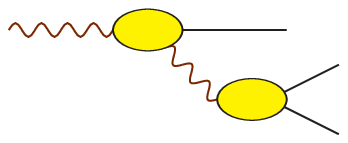,width=10cm}
\end{center}

\caption{\label{fig1}
Reducible one photon exchange 
diagram contributing to $F^{3\pi}(s,\cos\theta)$ via the 
electromagnetic form factor of the pion and the anomalous $\pi^0 \rightarrow
\gamma\gamma^*$ form factor $F^{\pi}(0,t)$.}

\indent
\end{figure}

\begin{figure}[t]
\begin{center}
\epsfig{file=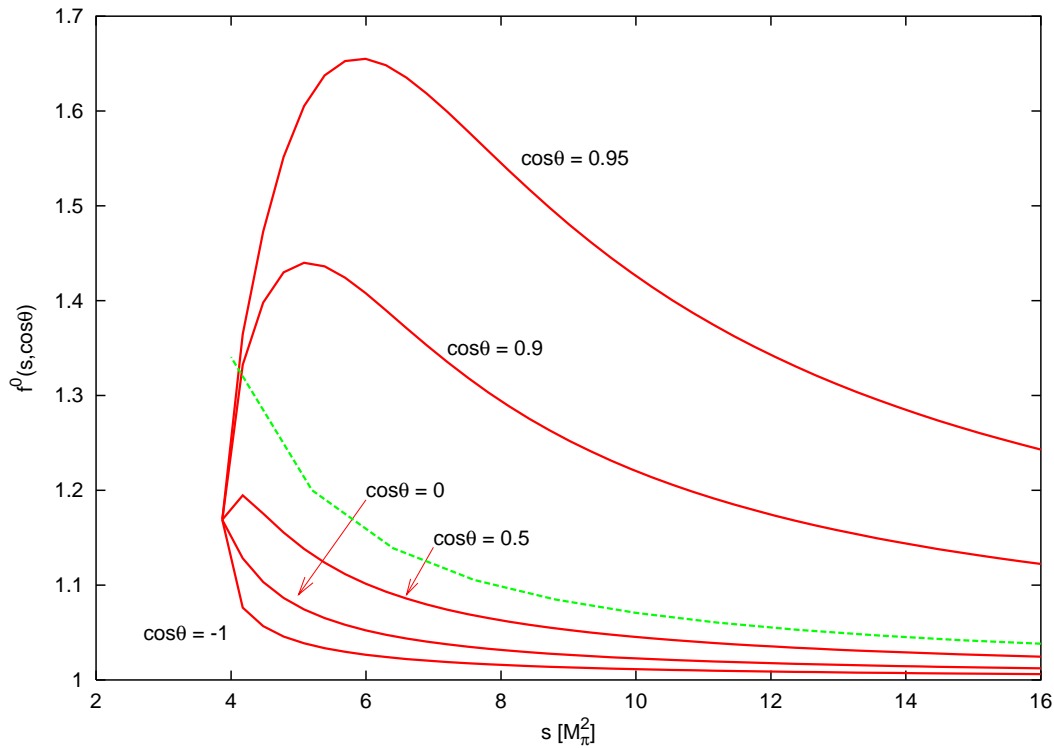,width=10cm,angle=-90}
\end{center}
\caption{\label{fig.LO}The lowest order amplitude $f^{(0)}(s,\cos\theta)$ 
including radiative corrections as a function of $s$ and for various 
values of the center-of-mass scattering angle $\theta$. The dashed curve 
shows the corresponding relative increase in the cross-section $\sigma$.}

\indent

\end{figure}

The second term, $f^{(1)}_{\mbox{\tiny FF}}(s,\cos\theta)$, 
arises from reducible diagrams with one photon
propagator as in Fig.~\ref{fig1}, where the blobs stand for form factors 
$F^{\pi}(0,t)$ and $F_V(t)$, computed at one-loop order,
\bea
f^{(1)}_{\mbox{\tiny FF}}(s,\cos\theta)
& =&
-2 e^2 \frac{1}{t}
\Big\{     \frac{1}{3} 
{\bar J}_{00}(t) (t-4M_{\pi^0}^2)
+\frac{t}{96\pi^2} \Big({\bar l}_6 -\ln\frac{M_{\pi^0}^2}{\mu^2}
-\frac{2}{3}\Big) 
\nonumber\\&&
-\frac{8}{3}\, F_\pi^2\Big( [A_2^r(\mu)-2 A_3^r(\mu)-4 A_4^r(\mu) ] 
M_{\pi^0}^2 
-[A_2^r(\mu)-4 A_3^r(\mu)] t\Big) 
\Big\}\,.
\label{f1FF}
\ena
Again, ${\cal O}(e^4)$ contributions have been neglected.
Notice that two kinds of counterterms contribute: 
${\bar l}_6$ has a non-anomalous origin and enters
through the pion electromagnetic form factor $F_V(t)$, 
whereas the terms  $A_i^r(\mu)$ belong to 
${\cal L}^-_{\rm \mbox{\tiny ep}^5}$. 
If we add this contribution to $f^{(0)}(s,\cos\theta)$, 
we obtain, as expected,
\eq
F^{3\pi}_0 [f^{(0)}(s,\cos\theta)\,+
\,f^{(1)}_{\mbox{\tiny FF}}(s,\cos\theta)]
\,=\,
F^{3\pi}_0\,-\,2e\,\frac{F^{\pi}(0,t)F_V(t)}{t}\,,
\label{poleterm}
\en
with the one-loop expressions of the form factors given as
\bea
F^{\pi}(0,t) &=& F^{\pi}_0\Big\{1 - 
\frac{8}{3}[A_2^r(\mu)-2 A_3^r(\mu)-4 A_4^r(\mu) ]M_{\pi}^2 
\nonumber\\
&& +
\frac{8}{3}[A_2^r(\mu)-4 A_3^r(\mu) - 
\frac{1}{256\pi^2}\ln\frac{M_{\pi^0}^2}{\mu^2}] t
+ \frac{1}{6F_\pi^2} 
{\bar J}_{00}(t) (t-4M_{\pi^0}^2) \Big\}\,,
\nonumber\\
F_V(t) &=& 1 + \frac{1}{6F_\pi^2} 
{\bar J}_{00}(t) (t-4M_{\pi^0}^2) +\frac{t}{96\pi^2F_\pi^2} 
\,({\bar l}_6-\frac{2}{3})\,.
\label{ffactors}
\ena

\begin{figure}[t]
\leavevmode
\begin{center}
\includegraphics[width=12cm]{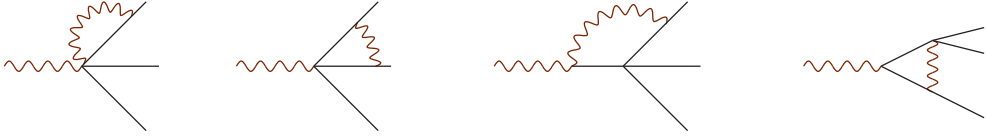}
\end{center}
\caption{\label{fig.IRD}
One-loop diagrams contributing to $\gamma \to
\pi \pi \pi$ via photon loops.}

\indent

\end{figure}

The last electromagnetic contribution, 
$f^{(1)}_{\mbox{\tiny IRD}}(s,\cos\theta)$,
comes from irreducible diagrams, shown in Fig.~\ref{fig.IRD}, 
with a virtual photon in the loop.
It is a ${\cal O}({\mbox{\rm e}}^2)$ correction to the tree-level result 
(the contribution from one-loop radiative correction to the 
$\gamma$-$\pi^+$-$\pi^-$ vertex of Fig. 1, is omitted, being  
${\cal O}({\mbox{\rm e}}^4)$)
and reads
\bea
\hspace{-0.7cm}
\hspace{-0.5cm}
f^{(1)}_{\mbox{\tiny IRD}}(s,\cos\theta) =&& 
e^2 
\Big\{ \frac{1}{16 \pi^2} \Big[ -2+ \frac{3M_{\pi^0}^2}{(s - M_{\pi^0}^2)}+
\frac{4 M_{\pi^0}^4}{(s-M_{\pi^0}^2)^2} - 
\log(\frac{m_\gamma^2}{M_{\pi^0}^2}) 
\nonumber\\&&\hspace{-0.5cm}
+\Big( \frac{3}{4} - 2 Z \Big) \ln\frac{M_{\pi^0}^2}{\mu^2} 
+\frac{Z}{6} ( {\bar k}_1-10 {\bar k}_2-3 {\bar k}_4 )
+\frac{3}{8} (3 {\bar k}_1-{\bar k}_3 ) \Big] + 
B_{\mbox{\tiny eff}}^r(\mu)
\nonumber\\&&\hspace{-0.5cm}
+{\bar J}_{0 0}(s) \Big[ 1
- \frac{M_{\pi^0}^2(M_{\pi^0}^2+s)}{ (M_{\pi^0}^2-s)^2} \Big]
+ {\bar J}_{\gamma 0}(s) \Big( 
\frac{M_{\pi^0}^2+s}{M_{\pi^0}^2-s} \Big)
\nonumber\\&&\hspace{-0.5cm}
+(\frac{2t-4 M_{\pi^0}^2}{t - 4 M_{\pi^0}^2}) \Big[
\frac{1}{8\pi^2}-{\bar J}_{0 0}(t)\Big] 
-(t-2 M_{\pi^0}^2) G_{+-\gamma}(t)
\nonumber\\&&\hspace{-0.5cm}
-2 M_{\pi^0}^2 H_{+-\gamma}(s;0)
+H_{+-\gamma}(0;s) \frac{2 M_{\pi^0}^6}{(M_{\pi^0}^2-s)^2}
\Big\}
\nonumber\\&&\hspace{-0.5cm}
+e^2
\Big\{ (s \leftrightarrow u) \Big\}\,.
\label{f1IRD}
\ena

\noindent 
Notice  the presence
of a non-vanishing photon mass $m_\gamma$. It is needed to regulate the 
infrared  divergence generated by the photon loop diagrams.

The functions $G_{+-0}(t)$ and $H_{+-0}(s;t)$ are related to the scalar 
three-point loop function $C_0$, defined as 
\eq
C_0(M_1^2,M_2^2,M_3^2;p_1^2,p_2^2,(p_1-p_2)^2) =
\frac{1}{i}\int \frac{d^d\,q}{(2\pi)^d} \frac{1}{(q^2-M_1^2) 
((q-p_1)^2-M_2^2) 
((q-p_2)^2-M_3^2)}\,,
\nonumber
\en
through
\bea
G_{+-0}(t) &=& 
C_0(m_{\gamma}^2,M_{\pi^\pm}^2,M_{\pi^\pm}^2;M_{\pi^\pm}^2,M_{\pi^\pm}^2,t)
\,,
\nonumber\\
H_{+-0}(s;t) &=& C_0(0,M_{\pi^\pm}^2,M_{\pi^\pm}^2;M_{\pi^\pm}^2,s,t)\,.
\ena
For $m_\gamma \to 0$, $G_{+-0}(t)$
can be expressed in terms of logarithms and dilogarithms. In the region of 
interest $t<0$, it is given by
\bea
&&G_{+-\gamma}(t) =
-\frac{1}{32 \pi^2 s \beta_t} \Big\{4 {\mbox Li}_2\left(\frac{1-\beta_t}
{1+\beta_t}\right)
+\ln^2 \left(\frac{\beta_t -1}{\beta_t +1} \right) +\frac{\pi^2}{3} 
\\ \nonumber&&
\hspace{3cm}
+2 \Big[ \ln\left(\frac{-t}{M_{\pi^\pm}^2}\right)-\ln\left(\frac{m_\gamma^2}
{M_{\pi^\pm}^2}\right)+2\ln(\beta_t) \Big]\,\ln\left(\frac{\beta_t -1}
{\beta_t +1}\right) \Big\}\,,
\ena
where 
$\beta_t\equiv \beta(t) = \sqrt{1-{4M_{\pi^\pm}^2}/{t}}$.

The infrared divergences
are handled as usual by considering the process with undetected soft 
photons, with  energies less than the detector resolution $\Delta E$. 
At the present level of accuracy, one soft photon is enough, and 
an infrared finite observable is constructed 
by considering in addition the cross-section $\sigma^\gamma(s;\Delta E)$
for the process $\gamma\pi^\pm\to\pi^0\pi^\pm\gamma$. Then, 
\eq
\frac{d\sigma+ d\sigma^\gamma}{\sin^3\theta d\theta}
\,\equiv\,(F^{3\pi}_0)^2\,\big\{
\vert f^{(0)}
 + f^{(1)}_{\mbox{\tiny QCD}} + f^{(2)_{\mbox{\tiny QCD}}} \vert^2
 \,+\, 2 {\mbox {Re} }[ f^{(1)}_{\mbox{\tiny IRD}}] 
 \,+\, 2 {\mbox {Re} }[ f^{(1)}_{\mbox{\tiny FF}}]\,\big\}
\,+\,
\frac{d\sigma^\gamma}{\sin^3\theta d\theta}\,,
\label{irfinite}
\en  
where we have also added the two-loop contribution 
$f^{(2)}_{\mbox{\tiny QCD}}$ 
{\it without} radiative corrections computed in Ref.~\cite{Hannah}, 
should be infrared finite.
The expression for the last term is cumbersome \cite{Van} and we do not
reproduce it here.  
We have checked that the infrared divergence appearing in 
the soft bremstrahlung term indeed cancels the one in 
$ 2 {\mbox {Re}} [{f}^{(1)}_{\mbox{\tiny IRD}}]$.
%
%

\subsection{Renormalization scale dependence of the counterterms}

In order to establish that $f^{(1)}(s,\cos\theta)$ indeed does not depend 
on the renormalization scale $\mu$, we need to know the scale dependence of 
the various counterterms involved in the expressions (\ref{f1QCD}), 
(\ref{f1FF}) and  (\ref{f1IRD}). This information has been obtained 
independently, using functional techniques, 
for the low-energy constants $l_i^r(\mu)$ \cite{gasser} 
and $k_i^r(\mu)$ \cite{marc_res,ulf}, but not for the remaining 
ones\footnote{The divergent part of the one-loop generating functional 
in the anomalous sector has actually been computed
\cite{anomaloop}, but the resulting 
expressions are rather cumbersome, and have, to the best of our knowledge, 
never been expressed in terms of the scale dependence of the 
renormalized constants $A_i^r$.}.
On the other hand, we may however use our present results 
in order to pin down the scale dependence of several 
combinations of the renormalized constants $A^r_i(\mu)$ 
and of $B^r_{\mbox{\tiny eff}}(\mu)$.
For instance, the form factor $F^{\pi}(0,t)$ (\ref{ffactors}), being 
an observable quantity, must be scale independent by itself, which implies
\bea
&&
\mu\,\frac{d}{d\mu}\,[A_2^r(\mu) - 2A_3^r(\mu) - 4A_4^r(\mu)] \,=\, 0\,,
\nonumber\\
&&
\mu\,\frac{d}{d\mu}\,[A_2^r(\mu) - 4A_3^r(\mu)] \,=\, 
-\,\frac{1}{128\pi^2F_{\pi}^2}\,.
\ena
Furthermore, the scale invariance of the amplitudes for 
$\gamma\gamma\to\pi^0\pi^0\pi^0$ and $\gamma\gamma\to\pi^+\pi^-\pi^0$ require
in addition that \cite{AKKT}
\eq
\mu\,\frac{d}{d\mu}\,
[A_7^r(\mu) - A_8^r(\mu) + 2A_{12}^r(\mu) - A_{13}^r(\mu) ] 
\,=\,
 -\,\frac{1}{128\pi^2F_{\pi}^2}\,.
\en
Then, the condition that $f^{(1)}(s,\cos\theta)$ does not depend on the 
renormalization scale $\mu$ amounts to
\bea
\mu\,\frac{d}{d\mu}\,[A_7^r(\mu) - A_8^r(\mu)] &=& 
 -\,\frac{3}{128\pi^2F_{\pi}^2}\,,
\nonumber\\
\mu\,\frac{d B^r_{\mbox{\tiny eff}}(\mu)}{d\mu} &=&
\frac{3}{32\pi^2}
\,.
\ena
Having discussed the scale dependence of the low-energy constants, it 
still remains to pin down their values in order to proceed.

\section{Counterterm estimates and numerical results}
\renewcommand{\theequation}{\arabic{section}.\arabic{equation}}
\setcounter{equation}{0}
\label{results}

In this section, we shall first provide estimates for the 
various counterterm combinations that appear in 
$f^{(1)}(s,\cos\theta)$. We then convert our computation into numerical 
results for the experimental observables discussed in section \ref{intro}.

\subsection{Counterterm estimates}
\label{cts}

The low-energy constant $\bar l_6$ contributes to the slope of the 
electromagnetic form factor of the pion \cite{gasser} 
and its value is well determined, 
$
\bar{l}_6 = 16.0 \pm 0.5 \pm 0.7\,
$
\cite{BCT}.
The combinations 
of constants $A_i$\footnote{The normalization of 
the low-energy constants $A_i$ we use here is the 
same as in Ref.~\cite{AKKT}, and differs from the one 
used in Refs.~\cite{fearing,KN} by a factor $-1/4\pi^2$.} 
which describe the ${\cal O}(M_{\pi}^2)$ corrections 
to $F^{\pi}(0,0)$ (\ref{ffactors})
and the slope of $F^{\pi}(0,t)$
were estimated by studying the $\langle VVP\rangle$ QCD 
three-point correlator \cite{bachir,AKKT,KN}. We shall use the 
estimates of Ref.~\cite{KN} (hereafter we set 
$\mu=M_{\rho}$) 
\bea
-\frac{8}{3} (A^r_2-2 A^r_3-4A^r_4) M_{\pi^0}^2  &=&
 (-6\pm 2 )\times 10^{-3}\,,
\nonumber\\
\frac{8}{3} (A^r_2-4 A^r_3) M_{\pi^0}^2 &=& (+26\pm 9)\times 10^{-3}
\,.
\label{pggcts}
\ena
A similar treatment of the combination of $A_i$'s appearing in 
$f^{(1)}_{\mbox{\tiny QCD}}$ would, for instance, require a detailed analysis 
of the $\langle VPPP\rangle$ four-point function along the lines of 
Ref.~\cite{KN}, which is lacking at the moment.
If we assume naive resonance saturation by the lightest
vectors, axials, pseudoscalars and scalars, only vectors contribute
to the QCD part. We obtain the rough 
estimate
\eq
- \Big\{ 8M_{\pi^0}^2 (2 A^r_{12}-A^r_{13}) -\frac{8}{3} 
(M_{\pi^0}^2+2M_{\pi^\pm}^2)
(A^r_7-A^r_8) \Big\} \sim \frac{2 M_{\pi^\pm}^2+M_{\pi^0}^2}{2 M_{\rho}^2}.
\en

The $\bar k_i$ constants appearing in  Eq.~(\ref{f1IRD}) are the finite, 
scale independent contributions of the electromagnetic counterterms $k_i$.
The precise relations among the non-renormalized $k_i$, the scale 
dependent renormalized $k_i^r(\mu)$
and the scale independent $\bar k_i$ have been worked out by several 
authors \cite{ulf,marc_res}. 
In \cite{marc_res}, the following estimates, based on  
naive dimensional analysis, were given
\be
\bar k_1=3.6 \pm 1.3 \,,\qquad \bar k_2= 3.5 \pm 1.3\,, \qquad
\bar k_3=3.4 \pm 0.7 \,,\qquad  \bar k_4=3.5 \pm 1.3\,.
\en

{F}inally, there remains the  effective anomalous electromagnetic 
counterterm $B_{\mbox{\tiny eff}}^r(\mu)$ in the anomalous sector, for which
naive dimensional analysis gives
\be
\vert\, B^r_{\mbox{\tiny eff}}(\mu)\vert\, \lapprox \frac{1}{16\pi^2} \,.
\en

\subsection{Numerical results}

Most of the existing results in the literature share a moderate 
increase of the amplitude $F^{3\pi}(s,\cos\theta)$
with $s$, and a very small
sensitivity to $\cos\theta$. 
This is, in particular, the case for the ChPT
results, both at one and two-loop order.  
When electromagnetic corrections are taken into account, 
two effects must be considered. 
First, a different mass for the neutral and charged pion modifies
the lower limit of the phase-space integral in Eq.~(\ref{data}). 
This represents small changes, an increase of 
$\sim  6\%$ in the Primakov cross section 
$\sigma^{\mbox{\tiny total}}/{\mbox Z}^2$, a
decrease of $\sim 4\%$ in  $F^{3\pi}_{0,\,{\mbox{\tiny exp}}}$, 
which goes into the right direction, but  still does not 
match the theoretical value.
Second, the radiatively corrected expression of  Eq.~(\ref{irfinite})
should be used in evaluating Eq.~(\ref{data}). 
In order to see how much the two combined effects modify 
the experimental determination of 
$F^{3\pi}_{0,\,{\mbox{\tiny exp}}}$, 
we show in Table \ref{table} the values obtained for $F^{3\pi}_0$
from the experimental result of Ref.~\cite{expe1}, using the  
expressions for the amplitude $F^{3\pi}(s,\cos\theta)$ 
at different orders of ChPT  in the $M_{\pi^0}=M_{\pi^\pm}$ isospin limit 
(first line),  and the corresponding 
values when ${\cal O}$(e$^2)$ electromagnetic 
corrections are included at one loop (second line).  
As numerical input parameters we have used \cite{PDG}
$
M_{\pi^\pm} = 139.570~\mbox{MeV}\,, 
M_{\pi^0} = 134.976~\mbox{MeV}\,, 
F_{\pi^\pm} = 92.4 ~\mbox{MeV}\,.
$
In addition, an energy resolution for undetected photons of 
$\Delta E=10$~MeV has been chosen.
We have checked that the final results are very insensitive 
to the actual value used for $\Delta E$ (we took values up to 50 MeV)
and for the next-to-leading order counterterms. The errors shown in 
Table \ref{table} have been computed adding the 
statistical and systematic experimental errors in Eq.~(\ref{expresult}).

\begin{figure}[t]
\begin{center}
\epsfig{file=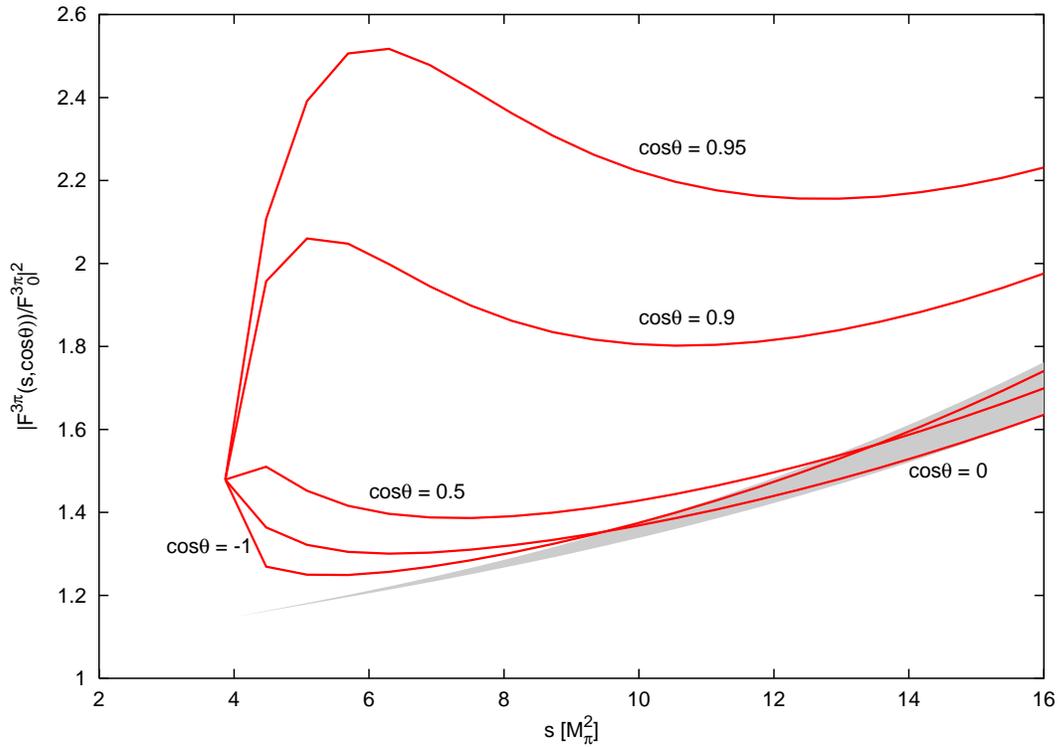,width=10cm,angle=-90}
\end{center}
\caption{
\label{superf} The scattering amplitude 
$\vert {F}^{3\pi}(s,\cos\theta)\vert^2$ vs. s (in units
of $M_\pi^2$) for several values of $\cos\theta$ at ${\cal O}$(p$^8)$ with 
(curves) and without (shaded area) 
radiative corrections. The shaded area covers the full range of  $\cos\theta$.
}
\end{figure}

It is interesting to observe that the electromagnetic 
corrections are sizeable:  
they represent an increase of $\sim 0.15$~nb
in the Primakov cross-section $\sigma^{\mbox{\tiny total}}/{\mbox Z}^2$, 
bringing the theoretical prediction to the value
\eq
\sigma^{\mbox{\tiny total}}_{\mbox{\tiny th}}/{\mbox Z}^2
\,=\,1.33\ \pm 0.03\ {\mbox{nb}}
\,,
\en
where the error has been estimated using higher order corrections 
via the non-perturbative 
chiral approach of Ref.~\cite{Hannah}. This value of the cross-section 
corresponds to 
\be
{\overline F}^{\,3\pi}_{\mbox{\tiny th}}
\,=\,11.7\ \pm 0.2\ {\mbox{GeV}}^{-3}
\en
for the average amplitude ${\overline F}^{\,3\pi}$.
Equivalently, if we keep $F^{3\pi}_0$ as a free parameter, and insert 
the theoretical expression (\ref{irfinite}) into Eqs. (\ref{data}) and 
(\ref{cross}), the experimental determination becomes
\be
F^{3\pi}_{0,\,{\mbox{\tiny exp}}}\,=\,10.7\pm 1.2\ {\mbox{GeV}}^{-3}\,,
\label{newF3piexp}
\en
much lower than the value (\ref{F3piexp}) obtained if 
radiative corrections are omitted, and compatible with the 
theoretical value $F^{3\pi}_{0}=(9.72\pm 0.09)\ {\mbox{GeV}}^{-3}$.
Interestingly enough, the contribution of the 
electromagnetic corrections to the amplitude $F^{3\pi}(s,\cos\theta)$
is larger than the genuine chiral corrections.  
This can be observed in Fig.~\ref{superf}, where
we plot the squared amplitude, as a function of $s$, for the 
${\cal O}(\rm\mbox{p}^8)$ result, 
including radiative corrections at one loop, for several  values of
$\cos\theta$ (curves). 
We also include a
shaded area for the same result without electromagnetic 
corrections, 
covering the full range of $\cos\theta$.
While the 
latter is almost 
insensitive to $\cos\theta$ and slowly raises with $s$, the
former 
shows a richer structure, with substantially large contributions for 
$\cos\theta$ approaching  $1$. 

This is in contrast with most other mesonic processes studied 
so far in ChPT, where electromagnetic
corrections in general give reasonably small contributions. 
The reason for this difference lies in a peculiarity of the 
$\gamma\pi^\pm\to\pi^0\pi^\pm$ process, which 
admits a kinematically enhanced contribution, due to the $t$-channel
exchange of a single photon, and
which actually completely dominates the radiative corrections.
As a matter of fact, our results can be very well reproduced by 
adding only the ``universal'' contribution coming from 
the pole in Fig.~\ref{fig1}\footnote{The residue of 
the pole is given to all orders in the strong 
interactions by $-2eF^{\pi}(0,0)F_V(0)$,
see Eq.~(\ref{poleterm}), 
which is equal to $-2eF^{\pi}_0$ to a very 
good approximation, see Eqs.~(\ref{ffactors}) 
and (\ref{pggcts}).} to the one and two-loop chiral corrections,
\eq
\frac{d\sigma}{\sin^3\theta d\theta}
\,=\,(F^{3\pi}_0)^2\,
\vert 1-2e^2\,\frac{F_{\pi}^2}{t}
 + f^{(1)}_{\mbox{\tiny QCD}} + f^{(2)}_{\mbox{\tiny QCD}} \vert^2
\,.
\en
The situation in this respect
would have been different in the case of the 
crossed  channel $\gamma\pi^0\to \pi^+\pi^-$. There, 
the photon-exchange
pole appears in the $s$-channel. With a minimum value
of $4 M_{\pi^\pm}^2$ for $s$, the net effect is mild, 
at the few-percent level.

\section{Conclusions}
\renewcommand{\theequation}{\arabic{section}.\arabic{equation}}
\setcounter{equation}{0}
\label{conclusions}

In the present work, we have considered,
within the framework of one-loop ChPT, 
${\cal O}(e^2)$ radiative corrections 
to the process $\gamma \pi^\pm\to \pi^0 \pi^\pm$.  
They turn out to be quite sizeable, being
larger than the genuine two-loop chiral corrections
computed in Ref.~\cite{Hannah}, and increase the two-loop
theoretical value for the Primakov 
cross-section  measured in Ref.~\cite{expe1} by $~\sim 0.15$~nb. 
Such a large effect originates mainly from a one-photon exchange contribution
in the $t$-channel. 
Since small kinematical values for $t$ are allowed, one obtains
large contributions to the cross-section.
Other electromagnetic contributions are very small.
These corrections are therefore very stable against variations of the
counterterms that enter at next-to-leading ${\cal O}$(e$^2$p$^2)$. 
The inclusion of radiative corrections brings theory and 
experiment into agreement. 
Future high precision experiments \cite{cebaf,compass,selex}
will hopefully improve
this agreement and test the r\^ole of electromagnetic 
corrections. In this respect, it would be worthwhile 
to investigate the experimental possibilities 
to study, in addition,
the reaction $\gamma\pi^0\to \pi^+\pi^-$, through {\it e.g.}
the process $\gamma p\to \pi^+\pi^- p$, where no such important 
electromagnetic effects are expected.

\vskip.4cm
\noindent {\bf Acknowledgments}\\
We thank A. Bramon, R. Escribano and J. Stern for 
clarifying discussions.
This work was supported in part 
by TMR, EC--Contract No. ERBFMRX--CT980169 (EURODAPHNE).

\newpage

\begin{table}[t]
\caption{
\label{table}Different chiral determinations of 
$F^{3\pi}_{0,\,{\mbox{\tiny exp}}}$
extracted from 
Eqs.~(\ref{data}) and (\ref{cross}),
as explained in the main text. The numbers 
in the first line  correspond 
to the  ${\cal O}(\rm\mbox{p}^4)$ tree-level, next-to-leading 
${\cal O}(\rm\mbox{p}^6)$
[9] and the dispersive treatment at ${\cal O}(\rm\mbox{p}^8)$ 
[10].
The second line correspond to 
the same results when electromagnetic corrections at  
${\cal O}(\rm\mbox{e}^2)$  
are included.
All results are in units of GeV$^{-3}$.
These values should be compared with the theoretical prediction 
$F^{3\pi}_{0}= 9.72\pm 0.09$ GeV$^{-3}$.
}
\begin{center}
\vspace{0.25cm}
\begin{tabular}{cccc}
& ${\cal O}(\rm\mbox{p}^4)$ & ${\cal O}(\rm\mbox{p}^6)$&
${\cal O}(\rm\mbox{p}^8)$ \\ 
$F^{3\pi}_{0,\,{\mbox{\tiny exp}}}(e=0)$ & 
$12.9\pm 1.4$ & $ 11.9 \pm 1.3$   
& $11.4 \pm 1.3$\\
$F^{3\pi}_{0,\,{\mbox{\tiny exp}}} (e\neq 0)$ & 
$12.0 \pm 1.2$ & $11.2 \pm 1.2$
& $10.7 \pm 1.2$  \\
\end{tabular}
\end{center}
\indent
\end{table}

\newpage

\newpage

\newpage

\newpage

\end{document}